# Abstractions for biomolecular computations.

Okunoye Babatunde O.

Department of Pure and Applied Biology, Ladoke Akintola University of Technology, P.M.B. 4000 Ogbomoso, Nigeria. babatundeokunoye@yahoo.co.uk

Deoxyribonucleic acid is increasingly been understood to be an informational macromolecule, capable of information processing. It has found application in the determination of non-deterministic algorithms and in the design of molecular computing devices. This is a theoretical analysis of the mathematical properties and relations of the molecules constituting DNA, which explains in part why DNA is a successful computing molecule.

Keyword: Computational complexity.

## Introduction.

DNA computing has become an established field within computer science; and several molecular computations and devices have been made[1,4,8,9], from operations on DNA molecules. A close study of the coding systems of DNA alerts us to the caveats of purely computational methods[5], the most obvious of which might be the property of degeneracy[2].

Our accumulated knowledge about biomolecular computing has not been matched by an understanding, in computer science terms, of the principles, which govern them[7]. By building on what we know about biomolecular systems based on the principles of molecular biology, we can create mathematical abstractions to explain why DNA molecules are accurate in information processing and computation. An abstraction – a mapping from a real-world domain into a mathematical domain – highlights essential properties while ignoring other, complicating ones[7]. Computer science can provide the much needed abstraction for biomolecular systems[7]. A good scientific abstraction has four properties: it is relevant, capturing the essential property of the phenomenon; computable, bringing to bear computational knowledge about the mathematical representation; understandable, offering a conceptual framework for thinking about the scientific domain; and extensible, allowing the capture of additional real properties in the same mathematical framework[7].

The bulk of DNA computations and devices show the capacity of DNA molecules to crunch algorithms and perform logical operations: a kind of miniature arithmetic and logical unit processor. Using the principles of molecular biology and mathematics, an abstraction or framework showing detailed algorithmic and logic operations encoded in DNA is proposed.

**DNA Algorithms**.

**The molecular structure of DNA is that of three bases coding for an amino acid[2]. In most cases, the third base in each codon plays a lesser role in specifying an amino acid than the first two, and in most cases, codons which specify the same amino acid differ only at the third base[6]. A definition of a program is given as program = algorithm + data structure[10]. By extension, algorithm = program – data structure. We bear in mind that DNA can be viewed as a high-level language or program, based on its properties and use as software in molecular computations. For the DNA molecule, data structures can be**

understood to be the amino acids specified for by triplet bases. Separating data structures from algorithms may be complicated in computer systems, but is quite uncomplicated in the molecule DNA. Deleting the third base from each codon thus obliterates data structures, leaving 16 doublet bases. These form the backbone of biomolecular algorithms.

The first DNA computation employed a 20 base-long oligonucleotide[1], as the basic algorithmic molecule. In keeping with DNA proportions, 20 is the closest multiple of ten to 16. There are ten bases per turn of the DNA helix[6].

The key to developing an algorithmic abstraction is by representing the doublet bases as numbers. The numbers of the base doublets were counted in segments of 20-base long oligonucleotides in 3,500 bases of Bacteriophage T4 Genome obtained from GenBank with accession number AF158101. The related complement is 168900' – 165,400', in the 5' to 3' direction. For each doublet, 175 numbers were obtained. The 'DNA – as – string' abstraction[7] is employed to create adjacency matrices. For each doublet base, the correlation coefficient of the first five numbers with the next five and so on is calculated. By using five numbers, we keep with DNA proportions: five numbers of base doublets correspond with ten bases, the length of a DNA helical unit. More importantly however, we make use of ten sets of numbers. A threshold correlation coefficient of ± 0.50 is fixed. Correlation coefficients in the range 0 - ± 0.49 is entered as zero, and correlation coefficients in the range ± 0.50 - ± 1.00 is entered as one. The result obtained under each doublet base is 35 0's and 1's, which represent entries into an adjacency matrix.

The correlation coefficient is one of the main tools in estimating the probability of an open reading frame (ORF) encoding a protein[5]. With slight modification above, we employ this tool to create adjacency matrices where '0' denotes 'no path' and '1' denotes 'a path' as in theoretical computer science.

This system creates a system of 16 × 16 adjacency matrices which can encode graphs of one to sixteen vertices. Leonard Adleman used a graph of seven vertices[1] to show the feasibility of DNA computing, so a 7 × 7 adjacency matrix is illustrated, showing one of the encoded Hamiltonian Paths encoded.

**DNA Logic.**

The key to understanding DNA logic is Kowalski's equation[3] : algorithm = logic + control. Rewriting this gives: logic = algorithm – control. Deleting the second base from each of the algorithmic base doublets removes control. The result is the 4 purine and pyrimidine bases. The language of DNA is composed of ten bases per turn of the helix[2]. by applying the 'DNA – as – string' abstraction, the numbers of these bases were recorded in 4,050 bases or 405 turns of the DNA helix. The related complement is 168,900' – 167101' in Bacteriophage T4 Genome, in the 5' to 3' direction. For each base, 405 numbers are obtained. Binary strings are created using the method of correlation coefficients as in algorithmic base doublets. In this case, DNA proportion is maintained by using ten numbers (ten bases).

By converting 0's to 'false' and 1's to 'true' since they are isomorphic, we obtain a truth table. Any of the four purine and pyrimidine bases can appear per ten bases (a turn of the helix), so we test the validity of the compound statement A í T í G í C ( A OR T OR G OR C). The resulting truth tables all show valid arguments (For Table 1: Since A í T í G í C are together true from lines 9 –16, the argument is valid. For Table 2: Since A í T í G í C are together true from lines 25 – 32, the argument is valid. For Table 3: Since A í T í G í C are together true from lines 33 – 37, 40 –44, the argument is valid. For Table 4: Since A í T í G í C are together true from lines 57 –64, the argument is valid. For Table 5: Since A í T í G í C are together true from lines 65 –72, the argument is valid.

**Conclusion.**

Biomolecular systems exist independently of our awareness or understanding of them, whereas computer systems exist because we understand, design and build them[7]. Nevertheless, the abstractions, tools and methods used to specify and study computer systems should illuminate our accumulated knowledge about biomolecular systems[7].

The surprising accuracy of molecular computations involving combinatorial problems and logic operations are caveats of other mathematical and theoretical relations within the molecules of DNA distinct from the basic relations: Adenine (A) complementary to Thymine (T), Guanine (G) complementary to Cytosine(C); $A + G = T + C$[6]. The author is not aware of any other work proposing the much needed abstractions for the emerging field of biomolecular computing.

**References.**

|    | AA | AT | AG | AC | TA | TT | TG |
|----|----|----|----|----|----|----|----|
|    | 1  | 0  | 0  | 1  | 0  | 0  | 0  |
|    | 1  | 1  | 0  | 0  | 1  | 1  | 0  |
|    | 0  | 1  | 1  | 1  | 0  | 0  | 0  |
|    | 0  | 1  | 0  | 0  | 1  | 0  | 0  |
|    | 0  | 1  | 0  | 0  | 1  | 1  | 0  |
|    | 0  | 1  | 1  | 1  | 0  | 1  | 0  |
|    | 1  | 1  | 0  | 1  | 0  | 1  | 1  |

† Fig. 1: **Adjacency Matrix**

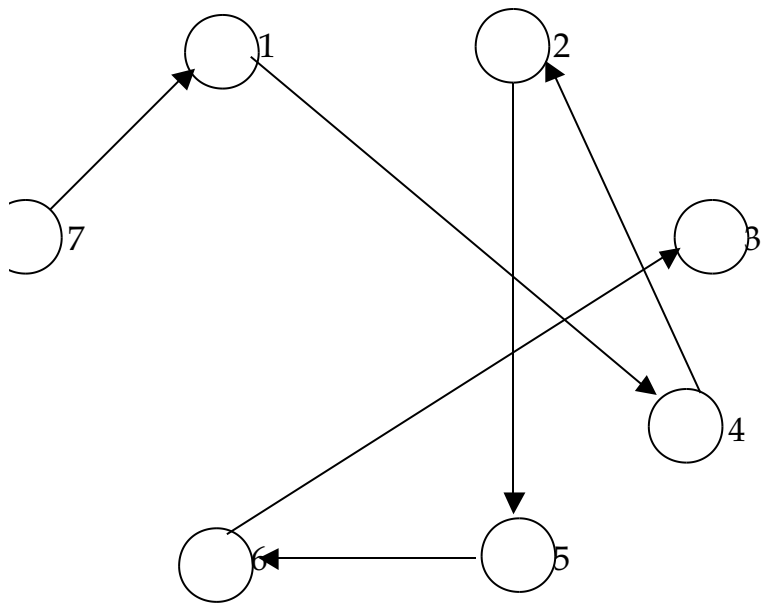

7 – 1 – 4 – 2 – 5 – 6 – 3

‡Fig.2: **Graph Algorithm.**

|    | A | T | G | C | A∨T | T∨G | G∨C | A∨C | A∨T∨G∨C |
|----|---|---|---|---|-----|-----|-----|-----|---------|
| 1  | F | F | F | F | F | F | F | F | F |
| 2  | F | T | F | F | T | T | F | F | T |
| 3  | F | F | F | T | F | F | T | T | T |
| 4  | T | F | T | T | T | T | T | T | T |
| 5  | F | T | F | T | T | T | T | T | T |
| 6  | F | F | F | F | F | F | F | F | F |
| 7  | T | T | T | T | T | T | T | T | T |
| 8  | F | T | F | F | T | T | F | F | T |
| 9  | F | T | T | F | T | T | T | F | T |
| 10 | F | F | T | T | F | T | T | T | T |
| 11 | F | F | T | T | F | T | T | T | T |
| 12 | T | T | F | F | T | T | F | T | T |
| 13 | T | T | F | T | T | T | T | T | T |
| 14 | T | F | F | T | T | F | T | T | T |
| 15 | F | T | F | T | T | T | T | T | T |
| 16 | T | T | T | T | T | T | T | T | T |
| 17 | T | T | F | F | T | T | F | T | T |
| 18 | F | T | F | T | T | T | T | T | T |
| 19 | F | F | F | F | F | F | F | F | F |
| 20 | F | F | F | F | F | F | F | F | F |
| 21 | T | F | F | F | T | F | F | T | T |
| 22 | F | F | T | F | F | T | T | F | T |
| 23 | F | F | F | F | F | F | F | F | F |
| 24 | T | F | T | T | T | T | T | T | T |
| 25 | T | T | T | T | F | T | T | T | T |
| 26 | F | T | F | F | T | T | F | F | T |
| 27 | T | T | F | F | T | T | F | T | T |
| 28 | F | F | T | T | F | T | T | T | T |
| 29 | F | T | F | F | T | T | F | F | T |
| 30 | T | F | F | F | T | F | F | T | T |
| 31 | T | T | F | F | T | T | F | T | T |
| 32 | F | T | T | F | T | T | T | F | T |
| 33 | T | F | F | F | T | F | F | F | T |
| 34 | T | T | F | F | T | T | F | T | T |
| 35 | T | T | F | F | T | T | F | T | T |
| 36 | F | T | T | F | T | T | T | F | T |
| 37 | T | F | T | F | T | T | T | T | T |
| 38 | T | T | F | F | T | T | F | T | T |
| 39 | F | F | F | F | F | F | F | F | F |
| 40 | F | F | F | T | F | T | T | T | T |
| 41 | F | F | T | F | F | F | T | F | T |
| 42 | T | F | F | F | T | F | F | T | T |
| 43 | F | F | F | T | F | F | T | T | T |
| 44 | F | F | T | T | F | T | T | T | T |

|    | A | T | G | C | A í T | T í G | G í C | A í C | A í T í G í C |
|----|---|---|---|---|-------|-------|-------|-------|---------------|
| 45 | F | F | F | F | F     | F     | F     | F     | F             |
| 46 | T | T | T | F | T     | T     | T     | T     | T             |
| 47 | F | T | T | T | T     | T     | T     | T     | T             |
| 48 | F | F | F | F | F     | F     | F     | F     | F             |
| 49 | F | T | F | T | T     | T     | T     | T     | T             |
| 50 | F | F | F | F | F     | F     | F     | F     | F             |
| 51 | F | F | F | T | F     | F     | T     | T     | T             |
| 52 | F | F | T | T | F     | T     | T     | T     | T             |
| 53 | F | F | T | T | F     | T     | T     | T     | T             |
| 54 | T | F | T | F | T     | T     | T     | T     | T             |
| 55 | T | F | F | F | T     | F     | F     | T     | T             |
| 56 | F | F | T | T | F     | T     | T     | T     | T             |
| 57 | T | F | F | T | T     | F     | T     | T     | T             |
| 58 | T | F | F | F | T     | F     | F     | T     | T             |
| 59 | F | F | T | T | F     | T     | T     | T     | T             |
| 60 | T | F | T | F | T     | T     | T     | T     | T             |
| 61 | F | T | F | F | T     | T     | F     | F     | T             |
| 62 | F | F | T | F | F     | T     | T     | F     | T             |
| 63 | T | F | T | F | T     | T     | T     | T     | T             |
| 64 | T | F | F | F | T     | F     | F     | T     | T             |
| 65 | T | F | F | F | T     | F     | F     | T     | T             |
| 66 | F | F | T | T | F     | T     | T     | T     | T             |
| 67 | F | F | T | F | F     | T     | T     | F     | T             |
| 68 | T | F | F | T | T     | T     | T     | T     | T             |
| 69 | F | T | F | T | T     | T     | T     | T     | T             |
| 70 | F | T | F | T | T     | T     | T     | T     | T             |
| 71 | T | F | T | F | T     | T     | T     | T     | T             |
| 72 | T | T | T | F | T     | T     | T     | T     | T             |
| 73 | T | T | F | T | T     | T     | T     | T     | T             |
| 74 | F | F | F | F | F     | F     | F     | F     | F             |
| 75 | T | F | T | F | T     | T     | T     | T     | T             |
| 76 | T | T | F | F | T     | T     | F     | F     | T             |
| 77 | F | T | F | F | T     | T     | F     | F     | T             |
| 78 | F | F | T | T | F     | T     | T     | T     | T             |
| 79 | F | T | F | F | T     | T     | F     | F     | T             |
| 80 | T | F | T | T | T     | T     | T     | T     | T             |